# Electronic Health Records: Cure-all or Chronic Condition?

Chris Kimble

**Abstract**

Computer-based information systems feature in almost every aspect of our lives, and yet most of us receive handwritten prescriptions when we visit our doctors and rely on paper-based medical records in our healthcare. Although electronic health record (EHR) systems have long been promoted as a cost-effective and efficient alternative to this situation, clear-cut evidence of their success has not been forthcoming. An examination of some of the underlying problems that prevent EHR systems from delivering the benefits that their proponents tout identifies four broad objectives - reducing cost, reducing errors, improving coordination and improving adherence to standards - and shows that they are not always met. The three possible causes for this failure to deliver involve problems with the codification of knowledge, group and tacit knowledge, and coordination and communication. There is, however, reason to be optimistic that EHR systems can fulfil a healthy part, if not all, of their potential.

## 1  Introduction

Computerization has touched almost every area of our lives. For decades, the worlds of commerce and manufacturing have invested heavily in computers and information technology and most of us now accept email, the Internet and mobile communications as part of everyday life. Yet despite this, most of us still receive handwritten prescriptions when we visit our doctors and rely on paper-based medical records in our healthcare. This is all the more surprising as, due in part to the increasing complexity of medical interventions and the increasing number of specialists, cost has become a major issue for healthcare provision in the industrialized nations. In addition, the trend has been toward shorter hospital stays combined with continuing treatment outside the hospital, with the resulting need to coordinate inpatient-type care in an outpatient setting. Industry-based information systems have been successfully dealing with the challenges of error reduction and cost reduction, along with the coordination of distributed workforces, for some time. Why has it not been possible for information technology to achieve similar results in healthcare?

Although the idea of using computers to process medical records is not new, it was not until the late 1980s and early 1990s that information systems began to deal with medical records on a routine basis. Systems built around clinical data, such as EMR (Electronic Medical Records), began to appear in the 1990s, followed at the turn of the century by such systems as EPR (electronic patient records) and EHR (electronic health records), which aim to provide comprehensive, cross-institutional, longitudinal records of patients' health and healthcare data. Despite the fact that similar systems had been used outside healthcare for a number of years, definitive evidence of their success in the healthcare domain remains elusive. Heeks, for example claims, "there are significant problems with significant numbers of information technology (IT) based systems in healthcare" (Heeks, 2006, p. 126). More recently, under the auspices of a European Union initiative, the EHR-IMPLEMENT program, the governments of Denmark, France and the United Kingdom have launched major projects to attempt to implement EHR in their countries. All of them failed (EU, 2012).

To determine the reasons behind the apparent inability to realize the benefits of IT in healthcare, it is helpful to look at what such systems are supposed to achieve and to analyse the reasons why these objectives are not being met. A number of detailed studies of the way such systems are used in practice present a picture of the persistence of paper-based documentation and the use of workarounds, such as letters and telephone calls, both of which appear to undermine the original rationale for the introduction of the systems. The theory underlying close industrial relatives, such as knowledge management systems and systems that support distributed collaborative work, helps identify three possible causes for this workaround behaviour.

## 2  A Review of the Expected Benefits of EHR Systems

One immediate problem with assessing the success or failure of systems like EHR is the board range of stakeholders involved and the range of diverse benefits that such systems are supposed to deliver. Greenhalgh et al (2009) for example, analysed 24 reviews of EHR studies and 92 primary studies to produce nine meta-narratives describing the different perspectives she found, whereas Hoerbst and Ammenwerth (2010) identified more than 1,200 quality requirements. As Heeks notes, "The first difficulty is the subjectivity of evaluation: Viewed from different perspectives, one person's failure may be another's success" (Heeks, 2006, p. 126).

The goal here is not to examine each expected outcome in detail, but simply to offer an overview that will provide a context for the analysis of the underlying issues that follows. The expected benefits of EHR can be classified under four broad headings: reducing cost, reducing errors, improving coordination, and improving adherence to a set of agreed standards.

### 2.1  Reducing the Cost of Healthcare

Cost reduction features prominently in the literature on health information systems from both Europe and America; however, the detailed analysis of the financial outcomes of such systems is equivocal (Greenhalgh et al., 2009). Most of the figures on the financial impact of EHR come from the United States, where the HITECH Act (part of the American Reinvestment and Recovery Act) has prompted many US hospitals to implement EHR systems or upgrade their existing systems. Similar developments in Europe also seem likely to lead to a pan-European system of electronic health records in the future.

The initial cost of EHR systems, which needs to be recouped by subsequent increases in revenue or cost savings, can be high. Menachemi and Collum (2011), for example, quote a figure of $19 million for a 280-bed, acute-care hospital. On the revenue side of the equation, EHR systems are purported to increase revenue by improving the tracking of resources and reducing billing errors, although this does not always appear to happen in practice (Monegain, 2013). In terms of cost savings, Linder et al. (2007) contend that many of the most widely cited figures in the United States appear to have been based on four benchmark hospitals that developed their own in-house EHR systems. Sidorov (2006) examines the claims made for potential cost savings in some detail. She notes that, for example, some of the productivity gains claimed for EHR could be realized only if the contracts for staff that it replaced were completely terminated but, in fact, many are redeployed to manage the new EHR system (Davidson & Chiasson, 2005). This is particularly significant as labour costs in American hospitals typically account for 40 to 50 percent of operating expenses.

### 2.2  Reducing Medical Errors

One of the first goals identified for medical informatics was the reduction of medical errors. Although Menachemi and Collum (2011) cite a number of examples that appear to show that this goal is being achieved, they also note that some studies seem to show the opposite. Sidorov (2006) looks at errors of omission (not doing something that should have been done), errors of commission (doing something that should not have been done), errors in medication, and the reduction in malpractice suits to reveal, once again, a mixed and apparently contradictory picture.

In their study of information system-related errors, Ash et al. (2004) help to shed some light on this by showing how such systems can sometimes create what they call "silent errors". Citing the example of the Therac-25 system, where the operators of a computer-controlled radiation therapy machine ignored obvious signs of tissue damage because they trusted the machine rather than their own observations, they suggest that physicians and nurses project a phantom objectivity onto the information produced by the systems, which they trust over their own instincts. The result is that such errors simply go unnoticed. More worrisome, they suggest that fixing these errors by adding new safety features can, in some situations, make things worse.

These findings are in line with research into safety critical systems, such as aircraft control systems, where each activity must follow another in a certain sequence. The analysis of accidents in these fields shows that problems tend to arise not from major errors, but from an accumulation of minor defects combined with a particular set of circumstances; consequently, the users of such systems must always be aware of the possibility of errors and continually seek to avoid them.

## 2.3   Improved Intra-organizational Coordination

As noted earlier, one of the problems with evaluating the benefits of EHR is the range of different stakeholders involved. It is, therefore, not surprising to find that there are differences in the approach to EHR between the United States and Europe. The European Union consists of 28 nations and has 24 official languages, plus a host of semi-official and minority languages. In addition, apart from a European health insurance card, which covers emergency medical treatment for the citizens from one European country when they visit another, a number of different systems are used to deliver healthcare at the national level. Consequently, there is a greater emphasis in Europe on EHR providing what Hoerbst and Ammenwerth (2010) describe as cross-border interoperability.

Nevertheless, one of the benefits claimed for EHR in both Europe and the United States is the ease with which medical records can be transferred among the various organizations involved with healthcare. This benefit is particularly obvious in the case of inpatient vs. outpatient treatment where a patient's records may need to be shared among several different medical bodies, as well as with organizations providing social support. Kindberg et al. (1999) provide an account of the complexities involved in this, whereas Takian, Sheikh, & Barber (2012) describe, in concrete terms, how EHR affected the work of the provision of mental health services in England. Their conclusion was that the systems improved the legibility and accessibility of patient records, and also provided insights into care processes, but did little to improve communication with the local social care services. Evidence from the United States is similarly uneven (Gittell & Weiss, 2004; Østerlund, 2004).

## 2.4   Improved Adherence to Standards

What is the role of EHR systems in improving the day-to-day working practices of healthcare personnel? In their study of the way in which an EHR system was developed, Davidson and Chiasson noted, "A major thrust of the project was to influence physicians' decision-making at the time clinical orders are issued, in ways that would reduce costs and improve patient care" (2005, p. 10). This is achieved by using the EHR system as a decision support system to prompt physicians to follow a particular, organizationally sanctioned clinical pathway in order to increase their compliance with agreed upon standards of best practice.

Winthereik et al. (2007), for example, looked at the role such systems can play in encouraging general practitioners in the United Kingdom to follow evidence-based rules in the management of chronic diseases, such as diabetes or heart disease, noting that, in general, doctors found it useful to be reminded about the steps they should take in the routine management of such conditions. Similarly, there is some evidence from the Unites States that EHR can help to improve certain aspects of preventive medicine, such as increasing the levels of vaccination against influenza (Menachemi & Collum, 2011) or referrals to smoking cessation clinics (Linder et al., 2007).

## 3   Looking at the Underlying Problems of EHR

Given the widespread use of IT in other sectors, what are the reasons behind the failure to realize the benefits of IT in health care? Before that question is addressed, it is helpful to look at the nature of the challenges faced by such systems in order to determine how EHR systems are used in practice.

### 3.1   Characterizing the Healthcare Environment

Even the briefest exploration of the topic of applying the types of information systems found in industry to healthcare will soon run up against the argument that healthcare is somehow fundamentally different (Gamble, 2013; Marcinko, 2011). Although it may have been possible to make such black-and-white distinctions in the past - when the sick, poor, and disadvantaged were housed and cared for in hospitals and those with mental illnesses were shut away in asylums - such comparisons in modern healthcare are far less clear-cut.

The organization of modern healthcare bears little resemblance to the institutions of the 19th and early 20th centuries, which were staffed by well-intentioned volunteers and run mainly by churches and charitable groups. Today, healthcare is provided by a wide range of specialized organizations, such as acute-care hospitals, dental practices, emergency treatment centres, general practitioners, home-based health services, hospices, nursing homes, psychiatric care facilities, rehabilitation centres, and walk-in clinics. In addition, healthcare is intimately connected to a host of other organizations not directly linked to its provision, such as clinical laboratories, employer-based healthcare plans, government agencies, insurance companies, medical equipment suppliers and manufacturers, pharmaceutical manufacturers, pharmacies, and regulatory bodies. Although it is no longer possible to view healthcare simply as a vocation, it is true to say that working in healthcare does provide some unique and challenging conditions that make the task of designing IT-based support particularly demanding.

Gittell and Weiss (2004) point to the problems of coordination and communication, both within hospitals and with outside agencies. They note that patient care presents a particular challenge, as its delivery is often highly fragmented, with diagnosis and treatment being supplied by a diverse group of loosely coupled service providers, even within a hospital setting where many of the resources have been brought together on a single site. Coiera (2000) and Ash et al. (2004), for example, stress that a great deal of the work carried out in hospitals is-event driven, with doctors and other healthcare professionals attempting to carry out several tasks while being constantly interrupted by beeper signals, telephone calls, and colleagues.

In addition to the problems of the working environment, Gittell and Weiss (2004) also point out that much of the medical information that needs to be exchanged is highly complex and not easy to codify, as is much of the information needed for the administration of healthcare. This problem is exacerbated in mental health settings, where long descriptive case notes are the norm (Takian et al., 2012) and may contain details of incidents not directly related to the patient's immediate care (Saario et al., 2012).

Patient records act as "boundary objects" that carry information from one professional group, such as doctors, to another, such as home care workers. If containers simply held unambiguous information, this would not pose a problem; however, the information contained in medical records is capable of being interpreted in a number of ways. The interpretive flexibility of patient records adds yet another layer of complexity to the problems of achieving effective communication and coordinating the range of activities that go into modern healthcare.

### 3.2  How Are EHR Systems Actually Used?

Thus, while the healthcare setting offers some particular challenges, these alone are not sufficient to explain why the success of information systems in healthcare appears to have been so variable. Understanding the reasons for this apparent lack of success requires a closer look at the way such systems are used in practice. An analysis of ethnographic studies can help. Ethnography is a method of data collection, originating in anthropology, which is now increasingly used in the design of computer systems and the study of technology. Its aim is to capture the social meaning of ordinary activities occurring in their natural setting. Typically, such studies involve extended periods of observing the way in which people carry out their daily activities. This involves taking detailed case notes, which are often supplemented by video and/or audio recordings, surveys, document searches, and interviews, to try to build up an understanding of whatever activity is being observed.

Ethnographic studies of EHR systems reveal the way such systems are used in reality, which may not be exactly what the designers or managers of the system had originally intended. Adopting this approach provides a much more nuanced evaluation of success and failure than can be obtained from headline cost/benefit figures. The ethnographic studies of EHR discussed below focus on settings that involve multiple agencies, multiple sites, or a number of different professions or specialties. They should help highlight the problems associated with the coordination of activities among different groups, which may be less obvious in studies of doctor-patient or doctor-computer interaction.

### 3.3  Ethnographic Studies of EHR in Multi-Agency Settings

The picture of EHR use found in most ethnographic studies is summarized in the title of an article by Saleem et al. (2011): "Paper Persistence, Workarounds and Communication Breakdowns". Although the common vision of an EHR system is that of an integrated information system that provides "a universal, patient-centred record which places a patient's entire medical history at doctors' and nurses' fingertips" (Østerlund, 2004, p. 35), the reality, it seems, is often far from this.

The continued use of paper in such systems is widely documented. From one perspective, this could be seen as merely a technical problem. For example, it is possible to argue that paper has certain advantages as a medium for storing and transmitting data over, say, a computer or a mobile device. Looking beyond the pros and cons of any particular medium, however, reveals a different picture. Saario et al (2012) describe two studies of what they term inter-professional electronic documents (I-PEDs), one in England and one in Finland. In both cases, they found that despite the fact that the I-PEDs were supposed to provide a standardized method for information exchange between different agencies, staff continued to make extensive use of alternative channels of communication, such as letters and telephone calls. Similarly, in their studies of the treatment of diabetic patients, Kindberg (1999) and Kindberg et al. (1999) found that paper notes were widely used. Even in so-called paperless practices, letters, telephone calls, and even the patients' own account of their treatment continued to be used as channels of communication. In these examples, the reasons went beyond simple convenience.

The problems of communication across professional boundaries in healthcare have been discussed elsewhere (Kimble, Grenier, & Goglio-Primard, 2010). The professionals in Saario et al.'s (2012) studies sometimes "censored" the information they entered into the system in order not to record hunches or details of events that might be relevant but were not directly related to the subject of the I-PED; letters or phone calls were used to communicate such information separately. Kindberg (1999) highlights the problem of deciding how the information contained in records should be interpreted. He notes that consultants are trained to be "sceptical investigators" and not to jump to conclusions. Each consultation consists of weighing up the evidence from a number of sources, not only on the progress of the condition, but also on the accuracy of what the patient has said and on the ability of others in the chain to undertake the various procedures that are effectively required. Consequently, the body of knowledge that the consultant deals with is not fixed, but is recreated at each consultation (Kindberg et al., 1999).

Although the circumvention of the concept of a single, comprehensive patient record by letters and phone calls may be understandable in the above examples, these are not the only circumstances in which other methods of exchanging information are used. Even in what would appear to be an ideal candidate for computerization, the planning of activities within a modern centralized hospital, multiple and apparently redundant sources of information continue to be used.

In his study of paediatric care, Østerlund (2004), for example, found several instances of the same information being recorded in multiple documents, some of which were being used concurrently. He details how a patient, Sophie, who had suffered a bad asthma attack, described her medical history 11 times to various doctors, nurses, clinical assistants, and administrators. He also describes how one of the people involved in Sophie's treatment, an intern, recorded information about the incident in four separate documents. Perhaps surprisingly in the light of what appears to be such a massive duplication of data, he concludes that his study questions whether the goal of a single, universal patient-centred record is the correct goal to follow. His argument is that the multiple records of what appears to be the same information are, in fact, not a record of what has happened to the patient, but an itinerary used by the various groups involved in a patient's care to coordinate their activities. He claims that although each group - interns, resident doctors, administrators, and so on - work in the same building with the same patients, the temporal and spatial patterns of their work do not overlap. He states that instead of a single patient-centred record, a work practice-centred record of activities is needed in order to coordinate patient care.

In their study of nursing handovers in a Norwegian hospital, Munkvold, Ellingsen, and Koksvik (2006) make similar observations. Ostensibly aimed at improving the documentation associated with handover procedures and formalizing the procedures themselves, here the system appeared to simply displace informality elsewhere. The researchers describe how, before the system was introduced, handovers took the form of stories about what had happened on the previous shift, involved the use of numerous documents, and would frequently overrun the allotted schedule. After the system was introduced, the meetings appeared to run on schedule and the nurses appeared to use the system to exchange information about the care of patients. In fact, Munkvold et al. found that the documentation produced by the system was printed out and heavily annotated by nurses for their day-to-day work, and that new examples of informal information exchange, such as weekly face-to-face summary meetings, began to appear. They concluded that the introduction of the system had, in fact, resulted in more undocumented informal work and information redundancy, although now it was largely hidden from the hospital's administrators.

Clearly some of the shortcomings of the current generation of EHR systems will be ameliorated by progress in technology and improvements in the design and implementation of such systems. The results from ethnographic studies, however, seem to imply that making systems such as EHR work will require more than technological enhancements and better change management.

## 4  Underlying Issues and Their Explanations in Theory

As noted earlier, the argument that healthcare was fundamentally different cannot be sustained in the modern world. Similarly, the problems associated with EHR do not appear to be simply a result of the shortcomings of the technology or poorly managed implementation. In fact, the underlying issues described above are not even unique to EHR; they have all been encountered, albeit to varying degrees, in other types of systems. The problems described below fall into three categories:

- The codification of knowledge
- Group and tacit knowledge
- Coordination and communication

The first two points are more often seen as an aspect of knowledge management, while the third is viewed as a problem of virtual or distributed team working.

### 4.1  The Codification of Knowledge

First, it is worth explicitly stating something that was alluded to in the previous section: Although medical records might accurately be described as persistent accumulations of data, they are not an unproblematical container for meaning. The data that they contain can be interpreted and used in a multitude of ways; the challenge for those who design and use such systems is to ensure that the data they contain are interpreted and used in the way that best serves the interests of the patients and those who treat them. This is a similar challenge to one that is faced in the design of information systems and knowledge management systems, where the solution has traditionally been to attempt to codify knowledge.

The idea behind the codification of knowledge is taken from Shannon and Weaver's (1949) mathematical model of communication. In their model, a message is encoded by a transmitter, which sends it through a communication channel to a receiver, where it is decoded to recreate the original message. As long as there is a stable shared code, the message will be transmitted accurately. Although Shannon and Weaver's model underlies much of the communication technology we use today, it is important to bear in mind that it is a mathematical model in which information is a quantity that can be measured in bits, and the semantic content of the message is irrelevant.

Notwithstanding this, the same broad approach can be applied to the messages stored in computers, such as records held in an EHR database. If the ways in which stored data can be interpreted can be restricted to meanings that have been allocated to a predefined, finite set of codes, then it is possible to avoid the problem of misinterpretation and to treat the data in much the same way as a message is treated in Shannon and Weaver's model. This approach is used widely in management information systems and forms the basis of many of what are often termed "first-generation" knowledge management systems.

Though effective in reducing the scope for misinterpretation, this approach is not without its cost, however. For example, Davenport and Glaser (2002) describe how Partners HealthCare developed a system that they say used technology to embed knowledge management into everyday work practices so that it was no longer seen as a separate activity - an undertaking they describe as both difficult and expensive. Similarly, Infosys, a business technology and IT services consultancy that has based much of its knowledge management strategy on this first-generation approach, found that it required a constant effort to maintain the "codebook" and to ensure that the relevant information was captured and codified (Kimble, 2013b).

### 4.2 Dealing With Group and Tacit Knowledge

Although codification is clearly an effective strategy, there is both a conceptual and a financial limit to codification. In some circumstances, codification is uneconomic, inappropriate, or simply undesirable (Kimble, 2013a).

Dealing with knowledge that needs to be shared by a group of people presents a particular problem, as each individual will have his or her own "codebook". In their article on the design of electronic communications systems to support knowledge work, Boland and Tenkasi (1995) make a direct comparison between Shannon and Weaver's model of communication and Wittgenstein's (1953) notion of language games. In contrast to Shannon and Weaver's shared and stable codebook, for Wittgenstein, the meanings given to words are unstable and change continually. According to Wittgenstein, the meaning of a message can be viewed only in the context of the constantly evolving knowledge and experiences of the person who receives it. Taken to the extreme, this would mean that each person who viewed a medical record would take away a different message.

In reality, of course, language is more often a way of connecting individuals by allowing them to express their internal experiences in a form that can be communicated to others. Meaning, in the sense of a working assumption of what others intend to convey, is not fixed, but is constructed on the fly through processes that allow participants to build, check, cross-check, and, if necessary, repair their understanding of what others are trying to convey.

The exchange of tacit knowledge poses a similar problem for approaches based on codification. Tacit knowledge is often described as knowledge that is difficult or impossible to put into words, which, as Gittell and Weiss (2004) point out, describes much of the information that needs to be exchanged in medical settings. If something cannot be articulated, however, neither can it be codified (Hildreth & Kimble, 2002). Again, the reality is that it is clearly possible to exchange knowledge without putting it into words. Groups that share a common set of beliefs, such as communities of practice and epistemic communities, create what is, in effect, an unstated and informal surrogate for Shannon and Weaver's formally defined codebook. Knowledge is shared through participation in activities based on a shared worldview, rather than by a formalized process of encoding and decoding messages.

In knowledge management, this view underlies what are sometimes termed "second-generation" systems. In healthcare, Coiera (2000) describes this as the "conversational" view of communication, which emphasizes social interaction and, like Kindberg's (1999) "sceptical investigators," sees the sharing and interpretation of information as an ongoing process of asking, telling, inquiring, and explaining, rather than a simple process of acquiring and retrieving data. For Coiera, systems such as EHR should not be viewed as information repositories, but as communication spaces where knowledge is co-created through repeated discussions and negotiations based on the data they contain. Shared values and beliefs, as well as shared professional backgrounds, act as the common ground for interpreting the data. These may be applied pre-emptively, as in the case of medical emergency teams who need to work together quickly and without misunderstanding, or in a just-in-time fashion when dealing with the exigencies of everyday work.

### 4.3 Dealing With the Coordination of Distributed Work

As Coiera's (2000) observations about medical emergency teams illustrate, the problems of coordinating actions within a group of individuals are closely related to the problems of sharing knowledge; in principle, the only difference is that the knowledge to be shared concerns what needs to be done and when, rather than the meaning of some particular term. One of the goals of systems such as EHR, however, is to reach out beyond the organization that originally collected the data and to improve inter-agency coordination by sharing that data with other healthcare providers. Coiera's notion of a conversation in which working assumptions can be checked, adjusted, and cross-checked might be appropriate for medical teams who are based in the same hospital and work together on a regular basis. It is less clear, however, how this model might be applied to people who work for different organizations, are located in different places, who may have never have met each other and who need to coordinate actions that may take place over a period of months or even years.

In contrast to the technological solutions found in other areas, the solutions to the problems of coordination in temporally and spatially distributed work tend to concern human relationships; most often, they are associated with dealing with the issues of trust and identity (Kimble, 2011). For example, the root cause of many of the workarounds that were documented in the multi-agency ethnographic studies were not the fault of the EHR system as such, but arose because the task that needed to be performed was spread across different groups that did not share the same level of trust and sense of common identity as Coiera's (2000) co-located co-workers. In this sense, Østerlund's (2004) call for a work-centred rather than patient-centred system was correct; to be effective these systems must not only support the medical needs of the patient, but they must also support the psychological needs of the people who look after them. Such support need not be high-tech. The workarounds found in the studies were all based on commonplace, standard technology. Sophisticated technologies, such as EHR, should not be seen as replacements for simpler technologies, such as letters, telephone calls, email, or shared directories, which still have a role to play and continue to be widely used.

## 5  The True Value of EHR

The frustration at the continued lack of success of EHR systems is succinctly expressed in the comments of a senior clinician, which became the title of Jones's study of the implementation of an electronic patient record system in a UK hospital: "Computers can land people on Mars, why can't they get them to work in a hospital?" (2003). So are electronic health records really a cure for the problems of modern healthcare, or just part of some chronic condition?

Research and experience seem to indicate that EHR systems are not, and will never be, a cure-all. A new technology does not simply replace whatever came before it. Photography did not replace painting, the cinema did not replace theatre, television did not replace the cinema, and it seems equally unlikely that EHR will completely replace the way in which healthcare records are administered at present. Heeks' (2006) observation that a failure viewed from one perspective may be seen as a success from another demonstrates that there are no straightforward solutions to how to improve the use of EHR systems. EHR may change certain aspects of healthcare delivery, but it is not the sole solution to improving it.

Codification is clearly a strategy that works. It is widely used in information systems and first-generation knowledge management systems, but it requires a continual effort to maintain the codebook and it is not always possible or desirable to do so (Kimble, 2013a). Similarly, evidence from the growth of Internet-based social networks shows that a second-generation approach to knowledge management is also viable. Going this route, however, means accepting that the fact that the "facts" contained in medical records can never be fixed, but will always remain fluid and open to interpretation and reinterpretation. The trade-off here is the ease of use and flexibility of Coiera's (2000) conversational model against the lack of ambiguity provided in the first-generation model.

Taken together, these give some reason for optimism for the future of EHR systems. Codification may not always be appropriate, and may also be costly to set up and maintain, but given these provisos, it does work. Similarly, as long as they are not viewed as problem-free universal repositories of information, systems based on communities and social networks can also play their part. The problem of distributed working, however, provides less room for optimism. Arguably, the difficulties associated with misunderstandings and a lack of trust of the motives of other groups, sometimes with good reason, has always been part of the human condition. Unlike with the previous two points, however, the solutions to this problem are primarily managerial rather than technological. Nonetheless, if the nature of the problem is recognized, then it should be possible to at least manage it.

If all the above points are taken into consideration, in time and with improvements in technology, the prognosis for the success of EHS systems is likely to improve, even though it appears that the underlying condition they are trying to treat will remain chronic.